\documentclass[pra,final,twocolumn,showpacs,showkeys]{revtex4}
\usepackage{amsmath}
\usepackage{graphicx}
\usepackage{amsfonts}
\usepackage{amssymb}
\usepackage{subfigure}

\newcommand{\subs}[1]
{ 
	\mbox{\scriptsize{#1}}
}

\begin{document}

\title{Higher-Dimensional Bell Inequalities with Noisy Qudits}

\author{Elena Polozova}
\author{Frederick W. Strauch}
\email[Electronic address: ]{frederick.w.strauch@williams.edu}
\affiliation{Williams College}

\date{\today}

\begin{abstract}
Generalizations of the classic Bell inequality to higher dimensional quantum systems known as qudits are reputed to exhibit a higher degree of robustness to noise, but such claims are based on one particular noise model.  We analyze the violation of the Collins-Gisin-Linden-Massar-Popescu inequality subject to more realistic noise sources and their scaling with dimension.  This analysis is inspired by potential Bell inequality experiments with superconducting resonator-based qudits.  We find that the robustness of the inequality to noise generally decreases with increasing qudit dimension.
\end{abstract} 
\maketitle

\section{Introduction}

A Bell inequality \cite{Brunner2014} experiment consists of $n$ parties who share an entangled state of $n$ (or more) particles.  Each party chooses to perform one of $m$ measurements, with each measurement producing one of $d$ outcomes.  By repeating this round multiple times, the probabilities for the various joint outcomes can be estimated.  A Bell inequality for this scenario is a relationship that these probabilities must satisfy, if they arise from a local realistic model.  The general structure of these inequalities has been studied intensively since Bell's original argument \cite{Bell65}.  

The most famous form of the Bell inequality is the Clauser-Horne-Shimony-Holt (CHSH) inequality \cite{Clauser69} for $n=2$ parties (e.g. Alice and Bob), each party performing one of $m=2$ measurement choices, with each measurement registering one of $d=2$ outcomes.  In terms of the joint probabilities, the CHSH inequality reads
\begin{equation}
p(A_1 = B_1) - p(A_1=B_2) + p(A_2=B_1) + p(A_2=B_2) \le 2,
\label{CHSH}
\end{equation}
where the measurement settings are labeled by $1$ and $2$ for Alice's (or Bob's) choice of measurement, with outcomes $A_1$ and $A_2$ for Alice (and $B_1$ and $B_2$ for Bob), and, in a slight abuse of notation, the joint probability $p(A_0=B_0)$ indicates the probability that Alice and Bob's measurement outcomes are identical.  For certain entangled states and measurement choices, quantum mechanics predicts, and experiments confirm, a violation of this inequality \cite{Brunner2014}.  

\begin{figure}
\includegraphics[width=3 in]{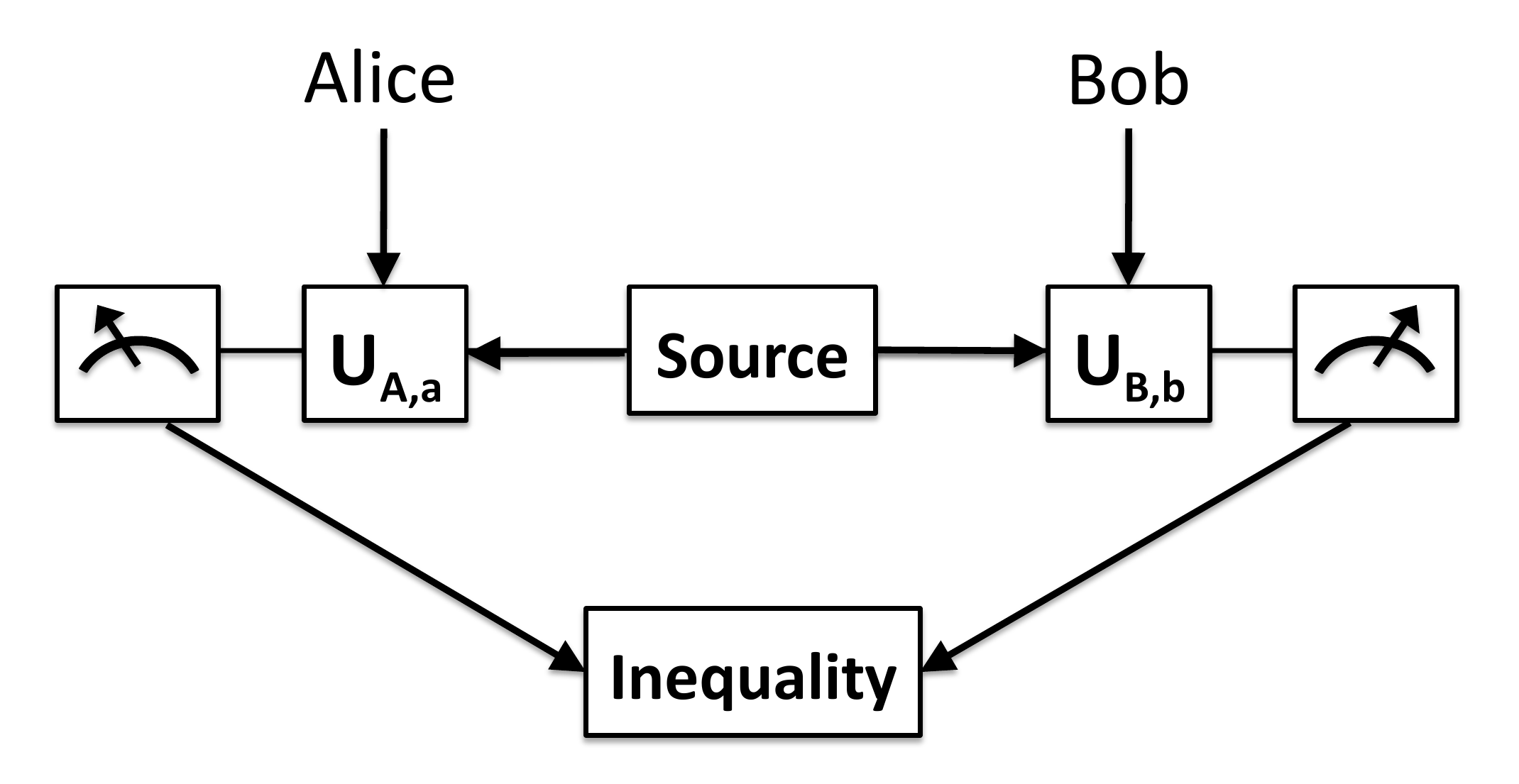}
\caption{A general framework for a Bell inequality experiment with $n=2$ parties.  A source of entanglement is shared to Alice and Bob, who choose to rotate their part of the joint quantum state by unitaries $U_{A,a}$ and $U_{B,b}$, respectively.  The measurement outcomes $M_A$ and $M_B$ are then recorded, compiled, and compared against the inequality.  }
\label{bellfig}
\end{figure}

A common experimental procedure, illustrated in Fig. \ref{bellfig}, replaces the alternative measurement settings by a unitary transformation chosen by the parties and performed just before a fixed measurement.  This unitary must be chosen and performed by one party sufficiently quickly to ensure that no information can propagate to the other party.   If this cannot be done, the experiment is subject to the so-called {\it locality} loophole, for which the derivation of Eq. (\ref{CHSH}) fails.  Another requirement is that the measurements be sufficiently accurate so that the probabilities entering the inequality can be reliably estimated.  If this is not the case, the experiment is subject to the so-called {\it detection} loophole.  Modern experiments with photons typically close the locality loophole \cite{Weihs98,Ursin2007}, but are subject to the detection loophole (in the guise of the fair-sampling assumption, but see recent progress \cite{Christensen2013,Giustina2013}).  Experiments with matter qubits (atoms \cite{Hofmann2012}, ions \cite{Rowe2001,Matsuk2008}, solid-state spin qubits \cite{Pfaff2013,Dehollain2015} and superconducting qubits \cite{Ansmann2009, Vlastakis2015}) typically close the detection loophole, but are subject to the locality loophole.   Achieving a loophole-free Bell inequality experiment would be a landmark test of quantum mechanics.

Towards this goal, and to better understand entanglement and nonlocality in general, there have been many studies of generalized Bell inequalites  \cite{Brunner2014}.  These include the Mermin inequality \cite{Mermin90} for multiple qubits ($n>2$), the Collins-Gisin inequality \cite{Collins2004} for multiple measurements ($m>2$), and the Collins-Gisin-Linden-Massar-Popescu (CGLMP) inequality \cite{Collins2002} for higher-dimensional systems known as qudits ($d>2$).  A recent approach \cite{Vertesi2010} to the Collins-Gisin inequality for entangled qudits (with  $n=2$ and $m=d>2$) exhibits the potential to reduce the requirements to close the detection loophole.

In this paper we focus on understanding the CGLMP inequality (for $n=m=2$ and $d>2$), which takes the form
\begin{equation}
I_d = \sum_{k=0}^{\lfloor d/2\rfloor -1} \left(1 - \frac{2k}{d-1} \right) \left[ \mathcal{P}(k) - \mathcal{P}(-k-1) \right] \le 2,
\end{equation}
where
\begin{eqnarray}
\mathcal{P}(k) &=& P(A_1 = B_1+k) + P(B_1 = A_2 + k + 1)  \nonumber \\
& & + P(A_2 = B_2 + k) + P(B_2 = A_1 + k).
\end{eqnarray}
Here the joint probabilities are defined for outcomes $A_a = 0, 1, \dots, d-1$, and the addition is performed modulo $d$.  These can expressed as
\begin{equation}
P(A_a = B_b + k) = \sum_{j=0}^{d-1} P(A_a = j, B_b = j + k \ \mbox{mod} \ d).
\end{equation}
We have also studied a closely related inequality proposed by Zohren and Gill \cite{Zohren2008}
\begin{eqnarray}
P(A_2 < B_2) + P(B_2 < A_1) + P(A_1 < B_1) & & \nonumber \\
+ P(B_1 \le A_2) &\ge& 1.
\end{eqnarray}
These two inequalities have the remarkable property that the violation increases with increasing $d$.  Alternatively, if the initial state is subject to depolarizing noise, the amount of noise that removes the entanglement, and hence the potential to violate the inequality, increases with the dimension \cite{Kasz2000}.  In this sense, these higher-dimensional Bell inequalities exhibit a surprising robustness to noise, and may be useful when exploring advanced tests of quantum systems.  

A recent experiment tested the CGMLP inequality using orbital angular momentum (OAM) states of light \cite{Dada2011}, but did not find the enhanced violation with dimension.  In fact, they found that the violation ceased for dimensions higher than $d=12$.  This fact strongly suggests that the robustness of the inequality can be compromised by the actual noise subject to the system.  Previous work on this topic has focused only on a simple form of depolarizing noise \cite{Kasz2000,Collins2002}, and its impact on the detector efficiency needed to observe a violation of the inequality \cite{Durt2001}.  To account for the behavior seen in the recent experiment, and to predict the possible violations in other experiments, more realistic noise models are necessary.  

In general, noise can affect all of the stages of the experiment: entangled-state preparation, state rotation, and state measurement.  The noise from each stage must be analyzed to determine the robustness of the inequality.   In this paper we examine a general framework for testing the CGLMP inequality with qudits, analyze the potential complexity of each stage of the experiment, and evaluate the impact of different types of noise (depolarizing, dephasing, and amplitude damping) on the inequality.  Our work is inspired by theoretical proposals to use the quantum states of superconducting resonators as qudits \cite{Strauch2010, Strauch11,  Krastanov2015}; such systems have the potential for long-distance entanglement through microwave photons \cite{Yin2013,Wenner2014}.  However, our analysis is intended to be general enough to guide experiments with other potential matter qudit systems, and our results may have implications for photonic qudits as well.  Our analysis indicates that the CGLMP inequality does not generally exhibit the special robustness to noise claimed in previous work.  

This paper is organized as follows.  Section II describes a general Bell inequality experiment with ideal qudits, analyzing  how state preparation, rotation, and measurement can be implemented for a general qudit system.  Section III analyzes the potential inequality violation subject to depolarizing, amplitude-damping, and dephasing noise on the qudits during the full experimental sequence.  In Section IV we conclude and discuss outstanding questions.    

\section{Bell Inequality Experiment with Ideal Qudits}

\begin{figure}
\includegraphics[width=3 in]{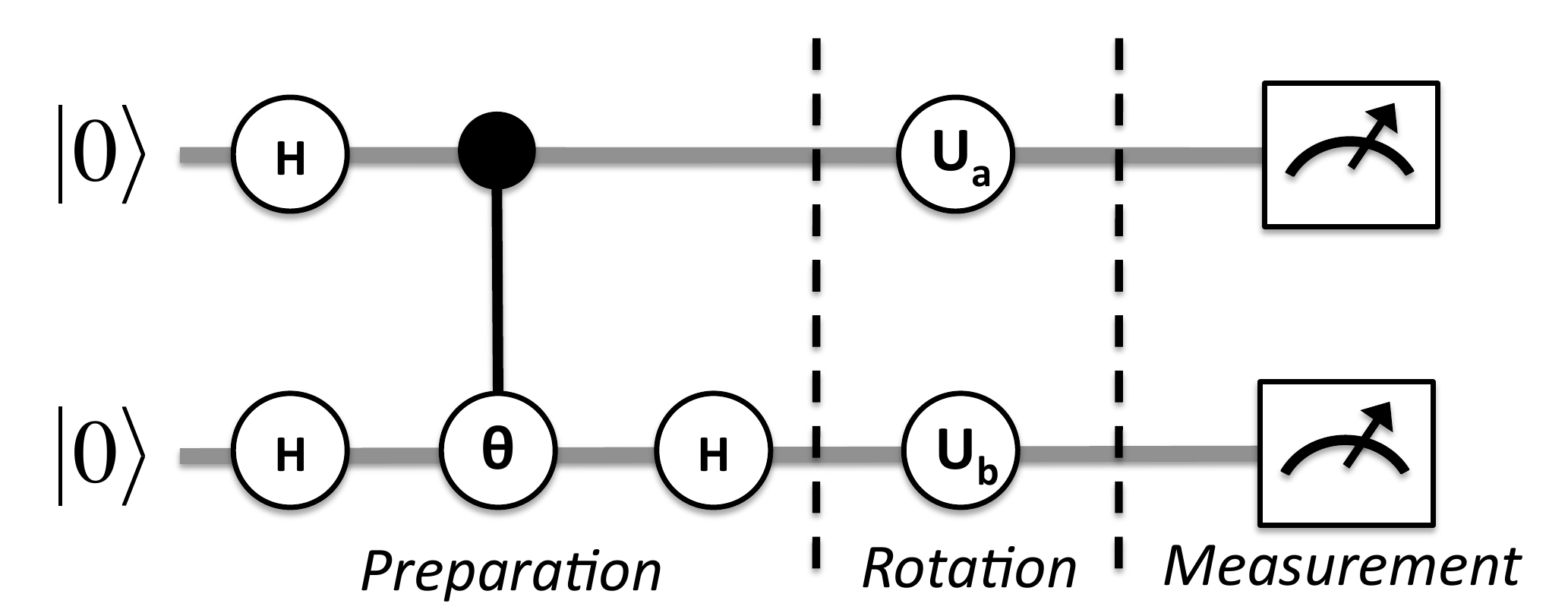}
\caption{A quantum circuit for a Bell inequality experiment, consisting of entangled-state preparation, single-qudit rotations, and qudit-state measurement.  In the first stage, two qudits, initially in the $|00\rangle$ state, are acted on by generalized Hadamard gates $H$, a controlled-phase gate (with $\theta = 2\pi /d$) to prepared a maximally entangled two-qudit state.  The single-qudit rotations $U_a$ and $U_b$ determine the two measurement bases (depending on the choices of Alice and Bob), while the actual measurement is in a fixed qudit basis. }
\label{bellcircuit}
\end{figure}

In this section we will describe an analysis of a bipartite ($n=2$) Bell inequality experiment, focusing on the general structure of entangled-state preparation, single-qudit rotations, and qudit-state measurement.  A quantum circuit for this process is shown in Fig. \ref{bellcircuit}.  We will analyze each component of this circuit, drawing inspiration from recent theoretical work on the control of superconducting resonators.  However, we expect our results can be applied to alternative qudit implementations.  

\subsection{Entangled-State Preparation}
The first important stage of a Bell inequality experiment is the production of entangled states of two systems.  As shown in Fig. \ref{bellcircuit}, this can be accomplished by using a generalized Hadamard or DFT (for Discrete-Fourier-Transform) operation on each qudit and a controlled-phase gate between the qudits.

The generalized Hadamard gate is the discrete Fourier transform (DFT) matrix with matrix elements $H_{j,k} = \omega^{jk}/\sqrt{d}$, where $\omega = e^{2\pi i /d}$.  For $d=4$, this has the matrix form
\begin{equation}
H(d=4) = DFT_4 = \frac{1}{\sqrt{4}} \left( \begin{array}{rrrr} 1 & 1 & 1 & 1 \\ 
1 & i & -1 & -i \\
1 & -1 & 1 & -1 \\
1 & -i & -1 & i
\end{array} \right).
\end{equation}  
In general this matrix has $d$ independent elements (and $d^2$ total elements), so an analysis of control on qudits suggests that implementing a general qudit Hadamard will require a number of elementary operations that is a polynomial in $d$ \cite{Brennen05}.

The controlled-phase gate is a two-qudit gate that implements the transformation
\begin{equation}
\mathcal{CR}(\theta)|j,k\rangle = e^{-i (j k) \theta} |j,k\rangle.
\end{equation}
This is a natural generalization of the two-qubit controlled phase gate, and can also be implemented in a time polynomial in $d$ \cite{Brennen05}.  

The combination of these operations in the preparation stage of Fig. \ref{bellcircuit} performs the transformation
\begin{eqnarray}
|\Psi\rangle &=& (H \otimes I) \mathcal{CR}(2\pi/d) (H \otimes H) |0,0\rangle \nonumber \\
&=& (I \otimes H) \mathcal{CR}(2\pi/d) \frac{1}{d} \sum_{j,k=0}^{d-1} |j,k\rangle \nonumber \\
&=& (I \otimes H) \frac{1}{d} \sum_{j,k=0}^{d-1} e^{-i 2 \pi (j k)/d} |j,k\rangle \nonumber \\
&=& \frac{1}{d^{3/2}} \sum_{j,k,\ell=0}^{d-1} e^{-i 2 \pi k (j-\ell)/d} |j,\ell\rangle \nonumber \\
&=& \frac{1}{\sqrt{d}} \sum_{j} |j,j\rangle.
\label{maxentstate}
\end{eqnarray} 
This maximally entangled two-qudit state will be used in the Bell inequality measurements for Alice and Bob.

We note that there are alternative approaches to generating the initially entangled state for superconducting resonators, which typically scale linearly in the qudit dimension \cite{Sharma2015}.  Qudit operations can also be optimized to scale linearly with qudit dimension \cite{OLeary06}.  We optimistically conclude that the state-preparation stage can be performed in a timescale that is linear in $d$.  

\subsection{Single-Qudit Rotations}

After Alice and Bob have chosen their measurement basis ($a$ or $b$), they will adjust their measurement by rotating their half of the joint-qudit state by one of the unitary operators $U_{A,a}$ or $U_{B,b}$.  These have the matrix elements
\begin{equation}
[U_{A,a}]_{j,k} = \frac{1}{\sqrt{d}} e^{ i 2 \pi (jk)/d} e^{i \alpha_a j}
\end{equation}
and
\begin{equation}
[U_{B,b}]_{j,k} = \frac{1}{\sqrt{d}} e^{-i 2 \pi (jk)/d} e^{i \beta_b j},
\end{equation}
where $\alpha_1 = 0$, $\alpha_2 = 1/2$, $\beta_1 = 1/4$, and $\beta_2 = -1/4$.  These operations involve the discrete Fourier transform (or its inverse), along with diagonal phases.  These again can be performed in a time polynomial in $d$, which we again optimistically take as linear in the qudit dimension \cite{OLeary06}.

\subsection{State Measurement}

In order to close both the locality and detection loopholes, one wants a fast and efficient measurement of the qudit states.  For some qudit implementations, such as hyperfine states of atoms or multilevel superconducting devices, one can implement a direct $d$ outcome measurement.   For qubit-resonator systems in trapped ion, cavity, or circuit-QED systems, one often implements an indirect measurement coupling the resonator to one or more qubits.   We consider a state measurement approach that maps the resonator state of $d=2^n$ dimensions onto $n$ qubits, inspired by the discussion in Chapter 6 of \cite{HarocheBook}.  After this mapping, one can measure the state bit by bit.   This allows a single-shot measurement of the qudit state.

\begin{figure*}
\includegraphics[width=6 in]{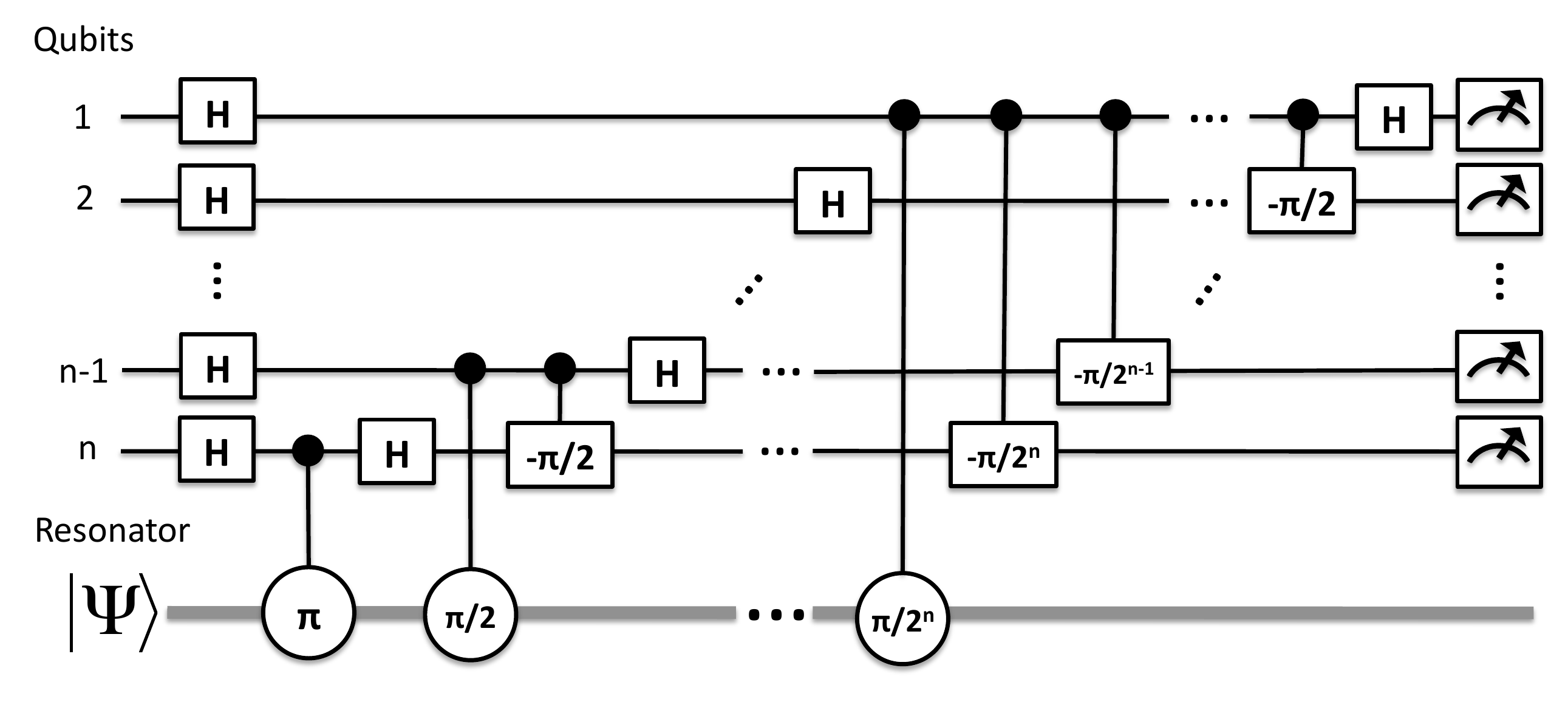}
\caption{A measurement circuit to map the quantum state of a resonator (of dimension $d = 2^n$) onto the state of $n$ qubits.  This circuit is composed of $2n$ single-qubit Hadamard operations,  $n$ qubit-resonator controlled-phase gates (with phases indicated in the circular gates on the resonator), and $n(n+1)/2$ qubit-qubit controlled-phase gates (with phases indicated in the rectangular gates on the qubits).}
\label{measurementfig}
\end{figure*}

The measurement scheme can be represented by a quantum circuit, illustrated in Fig. \ref{measurementfig}, composed of single-qubit Hadamard gates, two-qubit controlled-phase gates, and a special qubit-resonator controlled-phase gate.  This last gate acts on a qubit-resonator state as follows:
\begin{equation}
|x\rangle_{\subs{qubit}} \otimes |y\rangle_{\subs{res}}  \to e^{-i \theta x y} |x\rangle_{\subs{qubit}} \otimes |y\rangle_{\subs{res}}.
\end{equation}
This gate can be implemented using the natural evolution of a qubit-resonator system in the dispersive regime \cite{Strauch11}, resonant qubit-resonator logic operations \cite{Strauch2012}, or other methods \cite{Mischuck2013,Krastanov2015}.  In brief, the measurement circuit performs the mapping by sequentially correlating the resonator state $|y\rangle$ with a register of qubits ($\{x_1, x_2, \dots, x_n\}$).   Using a bit string $\{y_1, \dots, y_n\}$ to label the resonator state $|y\rangle$, with the (big-endian) binary representation
\begin{equation}
y = y_1 2^{n-1} + y_2 2^{n-2} + \cdots + y_{n-1} 2 + y_n,
\end{equation}
the circuit performs the mapping
\begin{equation}
|0 0 \cdots 0 \rangle_{\subs{qubits}} \otimes |y\rangle_{\subs{res}} \to |y_1 y_2 \cdots y_n \rangle_{\subs{qubits}} \otimes |y\rangle_{\subs{res}}.
\end{equation}
Then, measurement of the qubits will provide a bit-by-bit measurement of the resonator state $|y\rangle$.  

We now analyze the measurement circuit of Fig. \ref{measurementfig} in more detail.  The qubits are initially prepared in an equal superposition state (by Hadamard gates):
\begin{equation}
|\Psi_0\rangle = \frac{1}{2^{n/2}} \sum_{\{x_1, \dots, x_n\}, y} c_y |x_1 \cdots x_n \rangle \otimes |y \rangle
\end{equation}
The first qubit-resonator controlled-phase gate produces the state
\begin{equation}
\frac{1}{2^{n/2}} \sum_{\{x_1, \dots, x_n\}, y} c_y (-1)^{x_n y} |x_1 \cdots x_n \rangle \otimes |y\rangle.
\end{equation}
Using the binary representation, we have
\begin{eqnarray}
(-1)^{x_n y} &=& (-1)^{x_n y_1 2^{n-1}} (-1)^{x_n y_2 2^{n-2}} \cdots(-1)^{x_n y_n} \nonumber \\
&=& (-1)^{x_n y_n}.
\label{phaseeqn}
\end{eqnarray}
A subsequent Hadamard on qubit $n$ yields
\begin{equation}
\frac{1}{2^{(n+1)/2}} \sum_{\{x_1, \dots, x_n,z\},y} c_y (-1)^{x_n y_n + x_n z} |x_1 \cdots x_{n-1} z\rangle \otimes |y\rangle.
\end{equation}
Performing the sums over $x_n$ and $z$, using the fact that
\begin{equation}
\sum_{x_n=0}^1 (-1)^{x_n y_n + x_n z} = 2 \delta_{y_n,z},
\label{hadamardsum}
\end{equation}
we find
\begin{equation}
|\Psi_1\rangle = \frac{1}{2^{(n-1)/2}} \sum_{\{x_1, \dots, x_{n-1}\}, y} c_y |x_1 \cdots x_{n-1} y_n \rangle \otimes |y\rangle.
\end{equation}

The next qubit-resonator phase gate produces the controlled phase
\begin{equation}
\left(e^{i \pi/2}\right)^{x_{n-1}y} = \left(-1\right)^{x_{n-1} y_{n-1}} (i)^{x_{n-1} y_{n}},
\end{equation}
where we have used an argument similar to Eq. (\ref{phaseeqn}).  The final phase can be eliminated by a qubit-qubit phase gate, while a subsequent Hadamard on qubit $n-1$ yields
\begin{equation}
\frac{1}{2^{n/2}} \sum_{\{x_1, \dots, x_{n-1},z\},y} c_y (-1)^{x_{n-1} y_{n-1} + x_{n-1} z} |x_1 \cdots x_{n-2} z y_n \rangle \otimes |y\rangle.
\end{equation}
Performing the sums over $x_{n-1}$ and $z$, using the corresponding form of Eq. (\ref{hadamardsum}), the second stage yields
\begin{equation}
|\Psi_2\rangle = \frac{1}{2^{(n-2)/2}} \sum_{\{x_1, \dots, x_{n-2}\}, y} c_y |x_1 \cdots x_{n-2} y_{n-1} y_n \rangle \otimes |y\rangle.
\end{equation}

After $k$ stages we will find
\begin{eqnarray}
|\Psi_k\rangle &=& \frac{1}{2^{(n-k)/2}} \times \nonumber \\
& & \sum_{\{x_1, \dots, x_{n-k}\},y} c_y |x_1 \cdots x_{n-k} y_{n-k+1} \cdots y_n \rangle \otimes |y\rangle \nonumber \\
\end{eqnarray}
The next qubit-resonator gate produces the controlled phase
\begin{eqnarray}
\left(e^{i \pi/2^k}\right)^{x_{n-k}y} &=& \left(-1\right)^{x_{n-k} y_{n-k}} \times \nonumber \\
& &  \left(e^{i \pi/2}\right)^{x_{n-k} y_{n-k+1}} \cdots \left( e^{i \pi/2^k}\right)^{x_{n-k} y_n}. \nonumber \\
\end{eqnarray}
Eliminating all of the trailing controlled phases by qubit-qubit phase gates, performing a Hadamard gate on qubit $n-k$, and evaluating the summation over $x_{n-k}$ yields
\begin{eqnarray}
|\Psi_{k+1}\rangle &=& \frac{1}{2^{(n-k-1)/2}} \times \nonumber \\
& & \sum_{\{x_1, \dots, x_{n-k-1}\},y} c_y |x_1 \cdots x_{n-k-1} y_{n-k} \cdots y_n \rangle \otimes |y\rangle. \nonumber \\
\end{eqnarray}

Repeating until $k=n$, we finally obtain
\begin{equation}
|\Psi_n\rangle = \sum_y c_y |y_1 \cdots y_n \rangle \otimes |y\rangle.
\end{equation}
A measurement of the $n$ qubits will produce the outcome $(y_1, y_2, \dots, y_n)$ with probability $|c_y|^2$.  This circuit uses $n$ qubit-resonator gates, $2n$ Hadamard gates, and $n(n-1)/2$ qubit-qubit controlled-phase gates.  Thus, we see that this circuit requires a timescale of order $(\log_2 d)^2$. 

\section{Bell Inequality Experiment with Noisy Qudits}

The preceding analysis shows that, in general, one can expect that the state preparation, rotation, and measurement stages of a Bell inequality experiment will each scale with the qudit dimension.   If these stages are composed of fundamental elements, such as quantum gates, that are each subject to noise, then the resulting experiment will be subject to noise that scales with the qudit dimension.  How that scaling affects the inequality is the subject of this section.

While we will analyze idealized models of noisy qudits, our results can be understood in physical terms.  Each fundamental quantum gate will take some time, during which the quantum state can be subject to fluctuating fields, lose energy to the environment, or lose quantum information in some other fashion.  Each of these noise processes will affect the quantum state of the system, in terms of its density matrix.  If the total time for the experiment scales with the qudit dimension, the resulting density matrix will be subject to a correspondingly increased amount of noise.  

A detailed analysis of this process would require modeling the qudits' Hamiltonian and its coupling to external fields and the environment.  To understand the scaling with dimension, however, we can simplify our analysis to consider a single parameter figure-of-merit for each step of the experiment, such as the gate fidelity, and look at three types of noise: depolarizing noise, amplitude damping, and dephasing noise \cite{mikeandike}.  Using these models, we allow the strength of the noise to vary with the qudit dimension in the following way. 

We take the initial density matrix $\rho_0 = |\Psi\rangle \langle \Psi|$, with $|\Psi\rangle$ given by Eq. (\ref{maxentstate}), and iterate the appropriate trace-preserving map
\begin{equation}
\rho_{n+1} = \sum_{m} E_m \rho_n E_m^{\dagger},
\end{equation}
where the error operators $E_m$ depend on the type of noise, to be defined below, and are parametrized by a single number $p$.   This map is iterated $N$ times, where $N$ is proportional to the number of fundamental steps during the preparation and rotation stages of the Bell circuit of Fig. \ref{bellcircuit} and $p$ is proportional to the fidelity of each step.  The final density matrix, after $N$ applications of the noise map and subject to the measurement choices $a$ and $b$, is given by
\begin{equation}
\rho_{a,b} = (U_{A,a} \otimes U_{B,b}) \rho_N (U_{A,a} \otimes U_{B,b})^{\dagger}.
\end{equation}
The probabilities of the measurement outcomes are then
\begin{equation}
P(A_a=j, B_b=k) = \langle j,k| \rho_{a,b}|j,k\rangle.
\label{jointprob}
\end{equation}

As discussed above, the number of steps required to prepare the initial state and rotate the measurement is taken to be linear in $d$, while the single-shot measurement circuit scales with $(\log d)^2$.  For the dimensions $d \le 16$ considered below, there is little difference between linear and logarithmic scaling.  Thus, to understand how scaling affects the Bell parameter $I_d$, we set $N=d$.  For comparison, we will also consider $N=1$, similar to previous work.

\begin{figure*}[t]
\includegraphics[width=6 in]{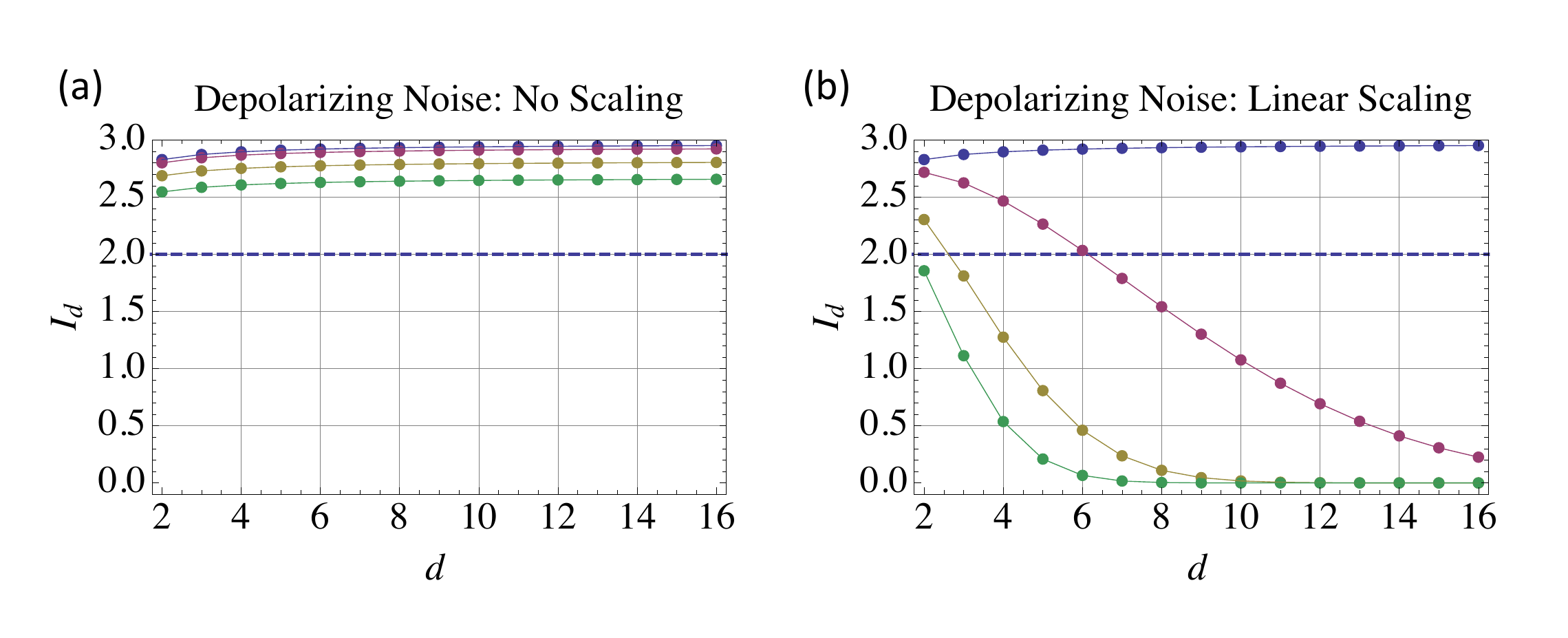}
\caption{Generalized Bell parameter $I_d$ as a function of qudit dimension $d$ for two types of depolarizing noise and various noise strengths.  (a) Single application of the depolarizing operator, with probabilities of $1$, $0.99$, $0.95$, and $0.9$, from top to bottom.   (b) Repeated application of the depolarizing operator with the number of applications is linear with the qudit dimension, with probabilities of $1$, $0.99$, $0.95$, and $0.9$, from top to bottom.   }
\label{depolarfig}
\end{figure*}

\subsection{Depolarizing Noise}

Depolarizing noise simulates the interaction of a system with a high temperature environment.  With each iteration, the system has probability $1-p$ of becoming depolarized, i.e. replaced by the completely mixed state, so that
\begin{equation}
\rho_{n+1} = p \rho_n +  (1-p) \frac{I}{d^2},
\label{depolar}
\end{equation}
while the system has probability $p$ of being unaffected.   Note that for this map, smaller values $p$ indicate stronger noise (this convention is chosen to match \cite{Collins2002}).  As described above, we allow this map to be repeated for a variable number of times before calculating the Bell parameter.  

The value of the generalized Bell parameter $I_d(p)$ is shown in Fig. \ref{depolarfig}, as a function of $d$ and for several values of the probability $p$.   From these results, we can make several observations.  First, for $p=1$, we see that the Bell parameter increases with dimension, as found previously.  In fact, we find
\begin{equation}
I_d(p=1) \approx 2.97 \left( 1 - \frac{1}{10 d} \right).
\label{bellincrease}
\end{equation}
Second, for $p < 1$ and $N=1$, the Bell parameter again increases with dimension, as seen in Fig. \ref{depolarfig}(a).  However, the overall violation does decrease with decreasing $p$.  In fact, it can be shown that, for this type of depolarizing noise, $I_d(p) = p I_d(1)$ \cite{Collins2002}.  This behavior underlies the reputed robustness to noise.

By constrast, if we apply the map $N=d$ times before calculating $I_d$ [Fig. \ref{depolarfig}(b)], we find that the violation decays with dimension.  This can be understood by noting that repeated iteration of Eq. (\ref{depolar}) has the solution
\begin{equation}
\rho_d = p^d \rho_0 + (1-p^d) \frac{I}{d^2},
\end{equation}
so that, in the presence of depolarizing noise scaling with $d$, the Bell parameter behaves as $p^d I_d(p=1)$.  Thus, the small increase in the Bell parameter (for $p=1$) of Eq. (\ref{bellincrease}) is quickly reduced by the exponentially decreasing factor $p^d$.  

This may have some bearing on the experimental results discussed above \cite{Dada2011}.  If one of the stages of the experiment has a fidelity that decays with the dimension of the entangled state, as evidenced in previous experiments \cite{Jack2009}, one can easily reproduce the observed decay of the inequality violation.  For example, setting $p=0.998$ and $N=d$ produces results in rough agreement with Fig. 3 of \cite{Dada2011}.

\subsection{Amplitude Damping Noise}

Amplitude damping describes the effects of energy dissipation on the quantum system.  We use a simplified amplitude damping model, chosen to represent the damping of a quantum resonator over a time $\Delta t$ with decay time $T$.  In this model, the singly excited state $|1\rangle$ decays with probability $p = e^{-\Delta t/T}$, while state $|j\rangle$ decays with probability $p^j$ \cite{Wang2008}.  

The specific model of amplitude damping is given by the map
\begin{equation}
\rho_{n+1} = \sum_{\ell,m=0}^1 (E_{\ell} \otimes E_m) \rho_{n} (E_{\ell} \otimes E_m)^\dagger,
\end{equation}
where the single-qudit amplitude-damping operators are given by
\begin{equation}
E_0 = \sum_{j=0}^{d-1} \sqrt{p^j} |j\rangle \langle j|
\end{equation}
and
\begin{equation}
E_1 = \sum_{j=1}^{d-1} \sqrt{1-p^j} |j-1\rangle \langle j|.
\end{equation}
We note that while this model intrinsically scales with dimension, we continue to allow the number of iterations of the map to scale with dimension as well.  It is also relevant to observe that this map is an approximation to the actual decay process, in that single application of this map with $p=0$ corresponds not to complete loss of energy, but to the subtraction of a single photon.  

\begin{figure*}
\includegraphics[width=6 in]{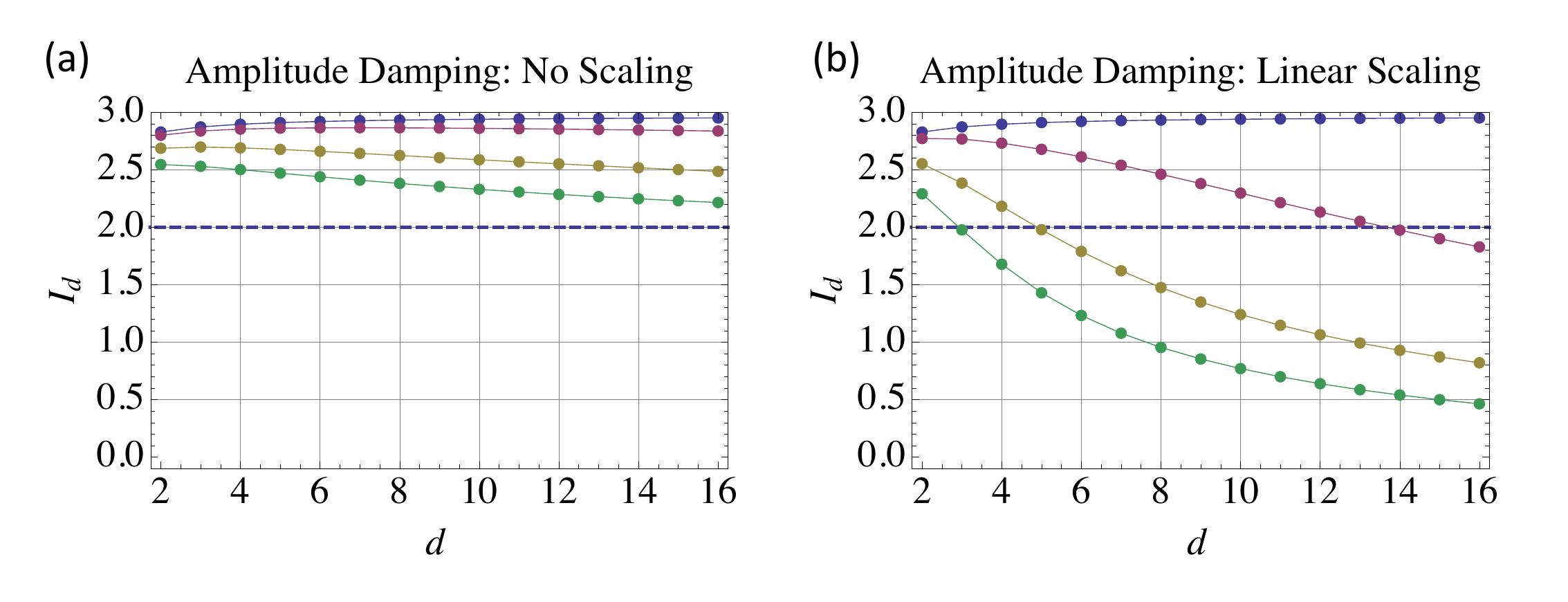}
\caption{Generalized Bell parameter $I_d$ as a function of qudit dimension $d$ for two types of amplitude damping and various noise strengths.  (a) Single application of the amplitude damping operators, with probabilities of $1$, $0.99$, $0.95$, and $0.9$, from top to bottom.   (b) Repeated application of the amplitude damping operators, with the number of applications is linear with the qudit dimension, with probabilities of $1$, $0.99$, $0.95$, and $0.9$, from top to bottom.   }
\label{ampdampfig}
\end{figure*}

Using this model of amplitude damping, the resulting Bell parameter $I_d$ as a function of dimension $d$ is shown in Fig. \ref{ampdampfig}, for several values of the probability $p$.  For $N=1$, the violation decays slowly with $d$, but remains a violation for up to $d=16$ for $p=0.9$.  Already, however, we see a significant difference from the behavior of $I_d$ under depolarizing noise.

For $N=d$ iterations, the violation decays much faster with $d$.  However, the decay is surprisingly less than the depolarizing map for the same value of $p$.  We will soon return to this issue, but for certain values of $d$ and choices of the probability, the two are similar.  For example, to model the OAM experiment \cite{Dada2011}, one could use amplitude damping with $p=0.992$ and $N=d$.     

The results above have used a discrete model of energy loss, in which a single quantum is removed from the system with some probability.  Of course, the actual physics involves a continuous loss of energy.  We have compared our discrete amplitude damping model to a more continuous model, by breaking up the decay into a number of intervals with the same overall decay factor $p$.  As one might expect, the difference between the two models is negligible for $p$ near 1, in which the probability of losing two quanta can be neglected.

\subsection{Dephasing Noise}
Dephasing describes loss of quantum information without loss of energy: rather than changing the amplitudes of the states as a function of time, the energy eigenstates of a system accrue random phases with some probability.   This is modeled by
\begin{equation}
\rho_{n+1} = p \rho_{n} + (1-p) \sum_{j,k=0}^{d-1} \left(\langle j, k| \rho_n |j, k\rangle \right) |j, k\rangle \langle j,k|.
\end{equation}
After each iteration, the off-diagonal elements of $\rho_n$ are reduced by a factor $p$.  

We note that the off-diagonal elements of the final density matrix in this case are identical to those found for the depolarizing map of Eq. (\ref{depolar}).  Remarkably, when measuring the density matrix using the DFT operators, the joint probabilities of Eq. (\ref{jointprob}) do not depend on the diagonal elements.  We thus arrive at the interesting result the effect of dephasing noise is identical to that of the depolarizing noise considered previously.  

To verify that the probabilities do not depend on the diagonal elements, we consider an arbitrary ``diagonal'' density matrix of the form
\begin{equation}
\rho_{\subs{diag}} = \sum_{j,k=1}^d c_{j,k} |j,k\rangle \langle j,k|,
\end{equation}
where
\begin{equation}
\sum_{j,k=1}^d c_{j,k} =1.
\end{equation}
For this density matrix we use Eq. (\ref{jointprob}) to calculate the joint probablitiies
\begin{eqnarray}
P(\ell,m) &=& \langle \ell,m| U_{A,a} \otimes U_{B,b} \rho_{\subs{diag}} \left(U_{A,a} \otimes U_{B,b} \right)^{\dagger} |\ell,m\rangle \nonumber \nonumber \\
&=& \sum_{j,k=1}^{d} c_{j,k} | \langle \ell | U_{A,a} |j \rangle|^2 |\langle m| U_{B,b} | k\rangle|^2 \nonumber \\
&=& \sum_{j,k=1}^{d} c_{j,k} \frac{1}{d^2} = \frac{1}{d^2}.
\end{eqnarray}
Since these probabilities are independent of $c_{j,k}$, the diagonal elements of any density matrix will contribute the constant value of $1/d^2$ to the probability $P(\ell,m)$.  Of course, the off-diagonal elements will also contribute to the probability.  However, any two density matrices with the same off-diagonal elements will lead to the same probabilities, confirming the result claimed above. 

We can also use this observation regarding the probabilities to help understand the difference between amplitude damping and depolarizing / dephasing noise.  For a single qubit, amplitude damping reduces the diagonal elements by $p$, but the off-diagonal elements by $\sqrt{p} > p$.  Thus, the effect of amplitude damping on the probabilities will be less than that for depolarizing / dephasing noise, for the same value of $p$.  That this holds true for dimensions $d>2$ is still somewhat surprising.  One possible explanation is that coherence between neighboring qudit states is largely maintained, much like the decay of a coherent oscillator state \cite{HarocheBook}, but we have not explored this conjecture further.  

\subsection{Thresholds for Inequality Violation}

A convenient way to summarize the results obtained above is to consider the minimum probability $p_{\subs{min}}$ for which the CGLMP inequality can be violated.  This probability satisfies
\begin{equation}
I_d(p_{\subs{min}}) = 2,
\end{equation}
and depends on the different models of noise.  Recall that a smaller probability $p$ indicates a greater amount of noise. For $p < p_{\subs{min}}$, the inequality is not violated, so we call $p_{\subs{min}}$ the threshold probability.  Previous work found that $p_{\subs{min}}$ decreases with dimensions, with the conclusion that higher dimensional Bell inequalities can be violated even if the system is subject to greater amounts of noise.  

\begin{figure}
\includegraphics[width=3 in]{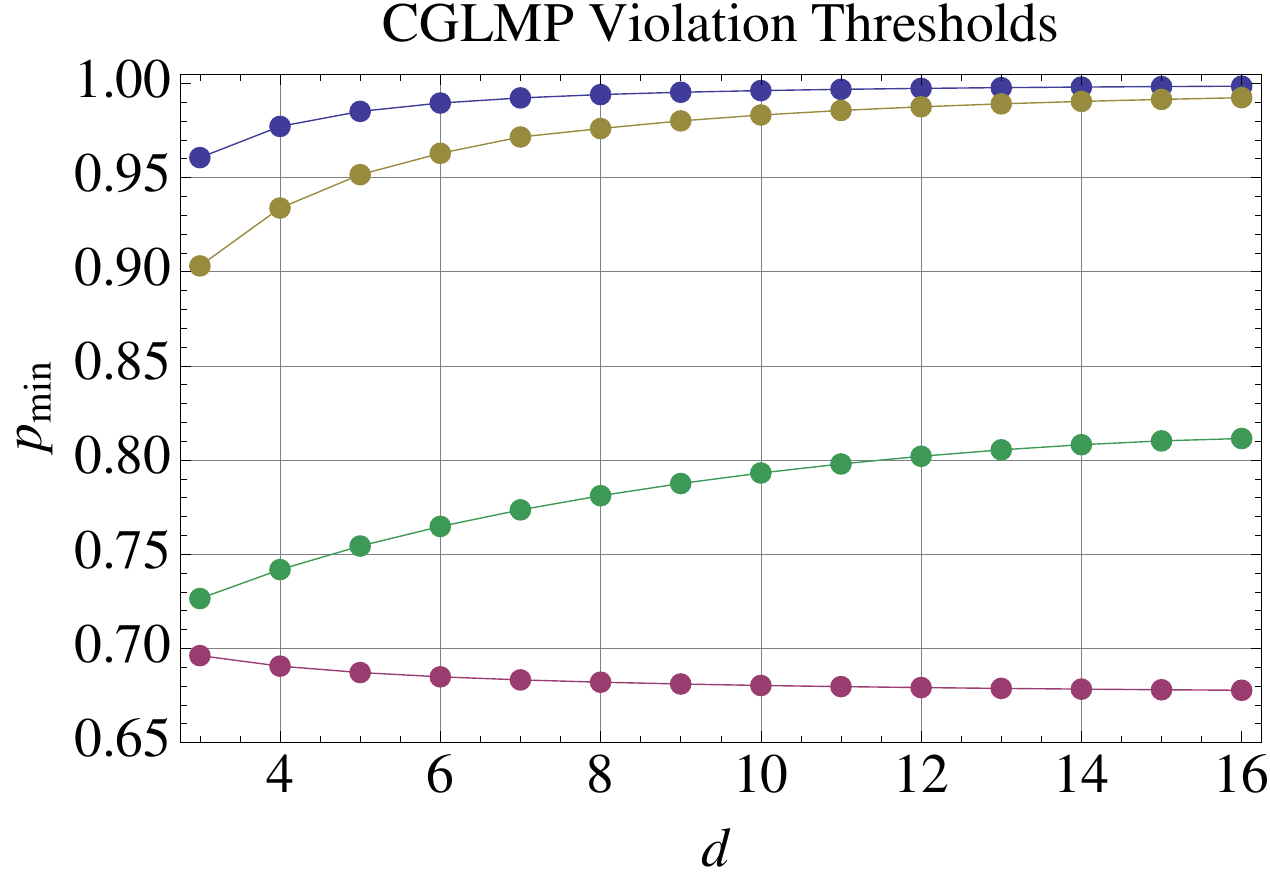}
\caption{The minimum probability $p_{\subs{min}}$ for which the CGLMP inequality is violated, as a function of qudit dimension $d$ and for the noise models (from top to bottom) depolarizing / dephasing noise with $N=d$, amplitude damping with $N=d$, amplitude damping with $N=1$, and depolarizing / dephasing noise with $N=1$.  Systems with noise $p < p_{\subs{min}}$ will not violate the inequality. }
\label{threshold}
\end{figure}

We have calculated the minimum probabilities for the various noise models described above and their dependence on dimension, as shown in Fig. \ref{threshold}.  We find that only for depolarizing / dephasing noise with $N=1$ does the threshold increase with dimension.  For all other cases---depolarizing / dephasing noise with $N=d$, and amplitude damping with $N=1$ and $N=d$---the minimum probability increases with dimension.  For these noise models, a violation of the CGLMP inequality with higher-dimensional systems requires higher-fidelity operations.  

\subsection{Alternative Procedures}

We have studied a number of alternative procedures of the Bell inequality test.  First, we have looked at the inequality proposed by Zohren and Gill \cite{Zohren2008} described in the introduction, and find that the threshold probabilities are identical to those shown in Fig. \ref{threshold}.  Second, we have looked at the alternative entangled states 
\begin{equation}
|\Psi\rangle_{\subs{app}} = \frac{1}{\mathcal{N}} \sum_{j=0}^{d-1} \frac{1}{\sqrt{(j+1)(d-j)}} |j,j\rangle,
\end{equation}
where $\mathcal{N}$ is a normalization factor; these states can achieve a higher degree of violation than the maximally entangled states \cite{Chen2006}, and approximate those states with maximal violation.  While the thresholds are slightly different for these states, they exhibit the same general behavior seen in Fig. \ref{threshold}.

Finally, we have considered another variation of the entangled state
\begin{equation}
|\Psi\rangle_{\subs{rev}} = \frac{1}{\sqrt{d}} \sum_{j=0}^{d-1} |j, d-j\rangle,
\end{equation}
in which Bob's qudit states have been reversed (Bob's unitaries $U_B,b$ must also be ``reversed'').  These states are easier to produce in superconducting circuits \cite{Sharma2015}, and one might think they would be less sensitive to amplitude damping.  However, the thresholds are again only slightly different for this alternative procedure.  In short, the behavior seen in Fig. \ref{threshold} appears to be generic for these inequalities and noise models.

\section{Conclusion}

We have studied the effects of noise on Bell inequality experiments using the higher-dimensional inequalites proposed by Collins, Gisin, Linden, Massar, and Popescu \cite{Collins2002} and Zohren and Gill \cite{Zohren2008} and variations thereof.  By modeling the required operations needed in the preparation, rotation, and single-shot measurement stages of these experiments, we have analyzed how the number of operations scale with the qudit dimension.  For most types of noisy operations, namely amplitude damping, depolarizing, and dephasing noise, we find that the higher-dimensional inequalities require increasingly higher-fidelity operations.  This conclusion runs counter to previous work, which considered depolarizing noise with a fixed noise probability, independent of qudit dimension.  However, this conclusion is in agreement with a very recent analysis of higher-dimensional Bell inequalities with random {\em coherent} errors \cite{Weiss2015}.

We conclude with a few observations.  Our results show that these higher-dimensional inequalities are not more robust against noise when taking into account the full experimental procedure.  Thus, these inequalities do not appear to offer a quick route towards closing the detection loophole with photons.  However, there are other multi-setting ($m>2$) inequalities with qudits \cite{Vertesi2010} that allow for reduced detection efficiencies for both atom-photon and photon-photon entanglement, and in the presence of depolarizing noise.  Our results motivate continued analysis of these and other inequalities (including those with $n>2$) in the presence of realistic noise appropriate for matter and photonic qudits.  The continued acquisition of evidence for nonlocality from a wide variety of physical systems remains an intriguing and important goal of modern quantum physics.

\acknowledgements
This work was supported by the NSF under Project Nos. PHY-1005571 and PHY-1212413.  

\bibliography{report}
\end{document}